\documentclass[aps,preprint,preprintnumbers,nofootinbib,showpacs]{revtex4}
\usepackage{graphicx,color,amsmath}

\newcommand{\newc}{\newcommand}
\newc{\be}{\begin{equation}}
\newc{\ee}{\end{equation}}
\newc{\beq}{\begin{eqnarray}}
\newc{\eeq}{\end{eqnarray}}

\begin{document}
\title{Neutrino number asymmetry and cosmological birefringence}
\author{C. Q. Geng$^{1,2}$,
S. H. Ho$^{1,2}$ and J. N. Ng$^{2}$}
\affiliation{$^{1}$Department of Physics, National Tsing Hua University, Hsinchu, Taiwan 300\\
$^{2}$Theory group, TRIUMF, 4004 Wesbrook Mall, Vancouver, B.C. V6T 2A3, Canada}
\date{\today}
\begin{abstract}
We  study a new type of effective interactions in terms of the
$CPT$-even dimension-six Chern-Simons-like term, which could originate from superstring theory, to generate
the cosmological birefringence.
We use the
neutrino number asymmetry to induce  a sizable
rotation polarization angle in the  data of the
cosmic microwave background radiation polarization.
The combined effect of the new term and the neutrino asymmetry
provides an alternative way to understand the birefringence.

\end{abstract}

\maketitle

The polarization maps of the cosmic microwave background (CMB)
have been  important tools for probing the epoch of the last
scattering directly. As we know, the polarization of the CMB can only
be generated by Thomson scattering at the last scattering surface
and therefore linearly polarized \citep{r-9,r-10}. When a
linearly polarized light travels through the Universe to Earth,
the angle of the polarization might be rotated by some localized
magnetized plasma of charged particles such as ions and electrons,
this is so-called Faraday effect. However, the rotated angle of
the polarization plane by this Faraday effect is proportional to the
square of the photon wavelength and thus it can be extracted.

On the other hand,
about ten years ago Nodland and Ralston \citep{r-1} claimed
that they found an additional rotation of synchrotron radiation
from the distant radio galaxies and quasars, which is
wavelength-independent and thus different
from  Faraday rotation, referred as the cosmological birefringence.
Unfortunately, it has been shown that there is no statistically significant signal present \cite{r-4,Comments}.
Nevertheless, this provides a new way to search for new physics in
cosmology.
Recently, Ni \cite{WTNi-pol} has pointed out that the change of the rotation angle of the polarization can be constrained at the
level of $10^{-1}$ by
 the data of
the Wilkinson Microwave Anisotropy Probe (WMAP) \cite{WMAP}
due to the correlation between the polarization and temperature.
 Feng $et\ al$ \citep{r-2} have used the combined data of the WMAP and
 the 2003 flight of BOOMERANG (B03) \cite{B03}
 for the CMB polarization to further constrain
the rotation angle  and concluded that
a nonzero angle is mildly favored. For a more general dynamical scalar, this rotation angle is
more constrained \cite{Liu}.
If such rotation angle does exist, it clearly
 indicates an
anisotropy of our Universe.
It is known that this phenomenon can be used to test the Einstein equivalence principle as was first pointed out by Ni \cite{WTNi,WTNi-r}.

Another theoretical origin of the birefringence was developed by Carroll, Field and Jackiw (CFJ)
 \cite{r-3}. They
modified the Maxwell Lagrangian by adding a Chern-Simons term \cite{r-3}:
\begin{eqnarray} \label{lagrangian}
\cal{L}&=&\cal{L}_{EM}+\cal{L}_{CS}
\nonumber \\
          &=&-\frac{1}{4}\sqrt{g}\emph{F}_{\mu\nu}\emph{F}^{\mu\nu}-\frac{1}{2}\sqrt{g}\emph{p}_{\mu}\emph{A}_{\nu}\emph{\~{F}}^{\mu\nu}\,,
\end{eqnarray}
where
$\emph{F}_{\mu\nu}=\partial_{\mu}A_{\nu}-\partial_{\nu}A_{\mu}$ is the electromagnetic tensor,
$\emph{\~{F}}^{\mu\nu}\equiv\frac{1}{2}\epsilon^{\mu\nu\rho\sigma}\emph{F}_{\rho\sigma}$
is the dual electromagnetic tensor, g is defined by
g=-det($g_{\mu \nu}$), and
 $p_{\nu}$ is a
four-vector.
 Here,  to describe a
flat, homogeneous and isotropic universe,
we use the Robertson-Walker metric
\beq
ds^2&=&-dt^2 +R^2(t)\;d\textbf{x}^2\,,
\eeq
 where R is the scale factor;
and the totally anti-symmetric tensor Livi-Civita tensor
$\epsilon^{\mu\nu\rho\sigma}=g^{-1/2}e^{\mu\nu\rho\sigma}$ with
the normalization of $e^{0123}=+1$.

 In the literature \citep{r-3,r-4,Carr1,Carr2,r-5,XZhang0611}, $p_{\mu}$ has been taken as a constant  vector or the gradient of a scalar.
In this paper, we study the possibility that the four-vector
$\emph{p}_{\mu}$ is related to a neutrino current
\beq
\emph{p}_{\mu}&=&\frac{\beta}{M^2}\emph{j}_{\mu}\,
\eeq
with the four-current
\beq
\emph{j}_{\mu}&=&\bar{\nu}\gamma_{\mu}\nu\;\equiv\;(j^{0}_{\nu}, \vec{\emph{j}_{\nu}})\,,
\label{nu-current}
\eeq
 where $\beta$ is the coupling constant of
order unity and M is an undetermined new physics mass scale.
Note that $\vec{\emph{j}_{\nu}}$
is the neutrino flux density and $j^{0}_{\nu}$ is the number density difference between neutrinos and anti-neutrinos, given by
\beq
j^{0}_{\nu}&=&\Delta n_{\nu}\;\equiv\; n_{\nu}-n_{\bar{\nu}}\,,
\label{Delta-n}
\eeq
where $n_{\nu(\bar{\nu})}$ represents the neutrino (anti-neutrino)
number density. It should be noted that if $\Delta n_{\nu}$ in Eq. (\ref{Delta-n}) is nonzero, the cosmological
birefringence occurs even in the standard model (SM) of particle interactions \cite{pal}. However, the effect is expected to be vanishingly small \citep{pal}.
In the following discussion, we will ignore this standard model effect.

As pointed out by CFJ
\citep{r-3}, in order to preserve
the gauge invariance we must require that the variation of
$\cal{L}_{CS}$, given by
\beq \cal{L}_{CS}&=&
-\frac{1}{2}\sqrt{g}{\beta\over
M^{2}}j_{\mu}\emph{A}_{\nu}\emph{\~{F}}^{\mu\nu}\,,
\label{Lcs}
\eeq
vanishes under the gauge transformation of $\Delta
A=\partial_{\nu}\chi $ for an arbitrary $\chi$.
However, one can check that in general, ${\cal L}_{CS}$ in
Eq. (\ref{Lcs}) may not be gauge invariant as
\begin{eqnarray} \label{CS}
\Delta\cal{L}_{CS}&=&\frac{\beta}{4M^{2}}\chi\emph{\~{F}}^{\mu\nu}(\nabla_{\nu}\emph{j}_{\mu}-\nabla_{\mu}\emph{j}_{\nu})
 \nonumber\\
 &=& \frac{\beta}{4M^{2}}\chi\emph{\~{F}}^{\mu\nu}(\partial_{\nu}\emph{j}_{\mu}-\partial_{\mu}\emph{j}_{\nu})\,,
\end{eqnarray}
which does not vanish generally.
To achieve the gauge invariance, one could use
the
St$\ddot{u}$ckelberg formalism\footnote{We thank Professor R. Jackiw for pointing out this possibility and an encouraging communication.}
\cite{jackiw}.
The Lagrangian in Eq. (1) can be reformulated by introducing one
St$\ddot{u}$ckelberg field $S^{\mu\nu}$
\begin{eqnarray}
\label{stuckelberg}
\cal{L}'&=&\cal{L}_{EM}+\cal{L}_{CS}'
\nonumber \\
          &=&-\frac{1}{4}\sqrt{g}\emph{F}_{\mu\nu}\emph{F}^{\mu\nu}-\frac{1}{2}\sqrt{g}\emph{j}_{\mu}
          (\emph{A}_{\nu}\emph{\~{F}}^{\mu\nu}+\partial_{\nu}S^{\mu\nu})\,,
\end{eqnarray}
where
 $S^{\mu\nu}$ is antisymmetric in
indices. It is clear that the requirement of the gauge invariance
is easily satisfied by acquiring a gauge transformation of
$S^{\mu\nu}$.
It is interesting to note that
${\cal L}'$ in Eq. (\ref{stuckelberg})
might originate
from the low energy effective theory in superstring theory\footnote{We are very grateful to Dr. W.F. Chen for showing us the string connection as well as sharing his
deep mathematical insights.}
in which the role of the St$\ddot{u}$ckelberg field
\cite{jackiw,S-more}
 is
played by the anti-symmetric Kalb-Ramond field $B_{\mu\nu}$
\cite{KR}.
For instance, by linking
$S^{\mu\beta}=\epsilon^{\mu\beta\sigma\rho}B_{\sigma\rho}$, it is straightforward to show \cite{chen} that our
effective interaction in Eq. (\ref{stuckelberg})
 has the same form as Eq. (13.1.42) in Ref. \cite{string}. We remark that the possible superstring origin for
 $\cal{L}_{CS}$ has also been given in Ref. \cite{Balaji} and the physical effects of the Kalb-Ramond field have been
 studied in Ref. \cite{Balaji,Kar}.

As we are working on the
usual Robertson-Walker metric,
the particle's phase space distribution function
is spatially homogeneous and isotropic, i.e. $\textit{f}
(p^{\mu},x^{\mu})$ reduces to $\textit{f} (\mid \vec{p}\mid,t)$
or  $\textit{f} (E,t)$  \citep{r-7}. In other words,
the relativistic neutrino background in our Universe is assumed
to be
homogeneous and isotropic like the CMB
radiation, which implies that
the number density for neutrinos is only a function of
red-shift z, i.e. the cosmic time. As a result, we conclude that the
neutrino current in Eq. (\ref{nu-current})
to a co-moving observer has the form
\begin{eqnarray} \label{current}
 \emph{j}_{\mu}&=&
\bigg{(}\Delta n_{\nu}\big{(}z(t)\big{)}, \vec{0}\bigg{)}\,.
\end{eqnarray}
Note that $\vec{j}=- D \vec{\nabla}\big[\Delta n_{\nu}\big{(}z(t)\big{)}\big]$,
 where D is diffusivity \citep{r-8} and $\vec{\nabla}$ is the
usual differential operators in Cartesian three-space. Here, we
have constrained ourselves to consider only the relativistic
neutrinos (for homogeneous and isotropic).

 From Eq.(\ref{current}), we have that
 $\partial_0 \emph{j}_i=0 $, $\partial_i
\emph{j}_0=\partial_i n_{\nu}(z)=0$ and $\partial_i \emph{j}_j=0$.
Consequently, we have a curl-free current for the co-moving frame.
%
%
In this frame,
the gauge invariance is maintained and
there is no need to include the St$\ddot{u}$ckelberg field.
However, the existence of a non-zero component $j_{\nu}^{0}$
 would violate Lorentz invariance \cite{r-3}.

It should be emphasized
 that the
Chern-Simons like term in Eq. (\ref{Lcs}) is $P$ and $C$ odd but
$CPT$ even due to the $C$-odd vector current of $j_{\mu}$ in
Eq. (\ref{nu-current}), whereas the original one
in Ref. \cite{r-3} is $CPT$-odd \cite{coleman}.
It is clear that ${\cal L}_{CS}$ in  Eq. (\ref{Lcs}) is a
dimension-6 operator and it must be suppressed by two powers of
the mass scale $M$.

Following
 Refs. \citep{r-4,r-3},
 the change  in the
position angle  of the polarization plane $\Delta\alpha$
 at redshift $z$
 due to our Chern-Simons-like term
  is given by
\begin{equation} \label{angle-1}
\Delta\alpha=\frac{1}{2}\frac{\beta}{M^2}\int \Delta n_{\nu}(t)
\frac{\textit{d}t}{R(t)}\,.
\end{equation}
To find out $\Delta\alpha$, we need to know the neutrino asymmetry
in our Universe, which is strongly constrained by the BBN
abundance of $^4$He. It is known that for a lepton flavor, the
asymmetry is given by: \citep{r-6,r-6r} \beq \label{asym-1}
\eta_{\ell}&=&\frac{n_{\ell}-n_{\bar{\ell}}}{n_{\gamma}}\;=\;\frac{1}{12\zeta(3)}
\left(\frac{T_{\ell}}{T_{\gamma}}\right)^3 (\pi^2
\xi_{\ell}+\xi_{\ell}^3)\,, \eeq where $n_{i}\ (i=\ell,\gamma)$
are the $\ell$ flavor lepton and photon number densities, $T_{i}$
are the corresponding temperatures and
$\xi_{\ell}\equiv\mu_{\ell}/T_{\ell}$ is the degeneracy parameter.

As shown
by Serpico and Raffelt \citep{r-6}, the lepton
asymmetry in our Universe resides in neutrinos because of the
charge neutrality, while the neutrino number asymmetry depends
only on the electron-neutrino degeneracy parameter $\xi_{\nu_{e}}$
since neutrinos reach approximate chemical equilibrium before BBN
\citep{r-11}. From Eq. (\ref{asym-1}), the neutrino number
asymmetry for a lightest and relativistic, say, electron neutrino
is then given by
\citep{r-6,r-6r,r-6more}:
\begin{equation} \label{asym-2}
\eta_{\nu_e}\simeq 0.249 \xi_{\nu_{e}}
\end{equation}
where we have assumed $(T_{\nu_{e}}/T_{\gamma})^3=4/11$.
Note that the current
bound on the degeneracy parameter is $-0.046<\xi_{\nu_{e}}<0.072$
for a $2\sigma$ range of the baryon asymmetry
\citep{r-6,r-6r}. From Eqs. (\ref{Delta-n}),  (\ref{asym-1}) and (\ref{asym-2}),
we obtain
\begin{eqnarray} \label{asym-3}
\Delta n_{\nu}&
         \simeq & 0.061\xi_{\nu_{e}}T_{\gamma}^3\,,
\end{eqnarray}
where we have used $n_{\gamma}=2\zeta(3)/\pi^2 \  T_{\gamma}^3$.
For a massless particle, after the decoupling, the evolution of its
temperature is given by
\citep{r-7}
\beq \label{temp-1}
T R &=& T_D R_D\,,
\eeq
where $T_D$ and $R_D$ are the temperature and scale factor at
 decoupling, respectively.
In particular, for $R=1$ at the present time, the photon temperature $T_{\gamma}^{\prime}$ of the red shift $z$  is
\beq
 \label{photon}
T_{\gamma}&=&\frac{T_D R_D}{R}=T_{\gamma}^{\prime}(1+z)\,.
\eeq
Then,  Eq. ({\ref{angle-1}) becomes
\begin{eqnarray}
 \label{angle-2}
\Delta\alpha &=&
\frac{\beta}{M^2} 0.030 \xi_{\nu_{e}} (T_{\gamma}^{\prime})^3 \int_{0}^{z_*} (1+z)^3
 \frac{\textit{d}z}{H(z)}\,,
\end{eqnarray}
where $H(z)$ is given by
\begin{equation} \label{H}
H(z)=H_0(1+z)^{3/2}
\end{equation}
in a flat and matter-dominated Universe and
 $H_0=2.1332 \times 10 ^{-42}h\ GeV$ is the Hubble constant
with  $h\simeq 0.7$ at the present.
Finally, by taking $1+z_*=(1+z)_{decoupling}\simeq 1100$ at the photon
decoupling and $T_{\gamma}^{\prime}=2.73K$,
we get \beq \label{angle-4} \Delta\alpha &\simeq& 4.2\times
10^{-2}\beta\left({\xi_{\nu_{e}}\over 0.001}\right)
\left({10\,TeV\over M}\right)^{2}\,. \eeq As an illustration, for
example, by taking $\beta\sim 1$, $M\sim 10\ TeV$ and
$\xi_{\nu_{e}}\sim \pm 10^{-3}$, we get $\Delta \alpha
\sim\pm4\times 10^{-2}$, which could explain the results in Ref.
\cite{r-2}. We note that a sizable $\Delta \alpha$ could be
still conceivable even if the neutrino asymmetry is small. In that case,
the scale parameter $M$ has to be smaller.

Finally, we note that there are several other
sources which can give rise to this cosmological
wavelength-independent birefringence. It is well-known the primordial gravitational vector or tensor perturbations in the CMB could produce a mixtrure of E-mode and B-mode polarizations and generate a non-zero rotation \cite{Seljak,Kamionkowski}. 
On the other hand, gravitational lensing also provides a source of the B-mode polarization of the CMB \cite{Zaldarriaga}. If there exists a cold neutral dark matter with a non-zero magnetic moment, it will serve as a source of the B-mode  CMB polarization and cause a non-zero wavelength-independent rotation angle \cite{Gardner}. In the presence of a quintessence background with a pseudoscalar coupling to electromagnetism, there can also be birefringence by the dynamical quintessence field \cite{Giovannini}.

In summary, we have proposed a new type of effective interactions
in terms of the
$CPT$-even dimension-six Chern-Simons-like term,
which could originate from superstring theory,
 to generate
the cosmological birefringence. 
To induce a sizable
rotation polarization angle in the CMB data,  a non-zero
neutrino number asymmetry is needed. 
We remark that
the Planck Surveyor \cite{Planck} will reach a sensitivity of $\Delta \alpha$
at levels of $10^{-2}-10^{-3}$ \cite{WTNi-r,Lue}, while a dedicated future experiment on
the cosmic microwave background radiation polarization would reach
$10^{-5}-10^{-6}$ $\Delta \alpha$-sensitivity \cite{WTNi-r}.

Note added: After the completion of this work, there was an
interesting paper by
Cabella, Natoli and Silk \cite{Silk}, which 
 applies a wavelet based estimator on the WIMAP3 TB and EB date 
 to constrain the cosmological birefringence.
 They derive a limit of $\Delta\alpha=-2.5\pm3.0$ deg, which is slightly
 tighter than that in Ref. \cite{r-2}.\\

\noindent
{\bf Acknowledgments}

We thank Prof. W.T. Ni, Dr. T.C. Yuan, Dr. Y.K. Hsiao and Dr. W. Liao for useful discussions.
This work is supported in part by
the Natural Science and Engineering Council of Canada
and
the National Science Council of
R.O.C. under Contract \#s: NSC-95-2112-M-007-059-MY3
and NSC-96-2918-I-007-010.

\end{document}